\documentclass[twocolumn,showpacs,preprintnumbers]{revtex4}
\usepackage{graphicx}
\usepackage{dcolumn}
\usepackage{bm}
\usepackage{epsfig}
\usepackage{amsmath}
\usepackage{subfigure}

\begin{document}

\title{Generating the Schr\"{o}dinger cat state in a nanomechanical resonator coupled
to a charge qubit}
\author{Jian-Qi Zhang$^{1}$}
\author{Wei Xiong$^{2}$}
\email{boing777@qq.com}
\author{Shuo Zhang$^3$}
\author{Yong Li$^4$}
\email{liyong@csrc.ac.cn}
\author{Mang Feng$^{1}$}
\email{mangfeng@wipm.ac.cn}
\address{$^1$State Key Laboratory of Magnetic Resonance and Atomic and Molecular Physics,
Wuhan Institute of Physics and Mathematics, Chinese Academy of Sciences, Wuhan 430071, China
$^2$Department of Physics and State of Key Laboratory of Surface Physics, Fudan University, Shanghai 200433,
China
\\ $^3$College of Science, National University of Defense Technology, Changsha
410073, China
\\ $^4$Beijing Computational Science Research Center, Beijing 100084, China}

\begin{abstract}
We propose a scheme for generating the Schr\"{o}dinger cat state based
on geometric operations by a nanomechanical resonator coupled
to a superconducting charge qubit. The charge qubit, driven  by two strong classical fields,
interacts with a high-frequency phonon mode of
the nanomechanical resonator. During the operation, the charge qubit
undergoes no real transitions, while the phonon mode of the
nanomechanical resonator is displaced along different paths in the phase
space, dependent on the states of the charge qubit. This generates the entangled cat 
state between the NAMR and charge qubit, and the cat state for the superposition 
of NAMR can be achieved after some operations applied on this entangled cat state. 
The robustness of the scheme is justified by considering noise from environment,
and the feasibility of the scheme is discussed.
\end{abstract}
\maketitle

\section{Introduction}

Recently, nanomechanical resonators (NAMRs) have attracted considerable
attention both theoretically and experimentally, due to the fact that NAMRs
own both classical and quantum properties \cite{prl-97-237201,Physics-2-40},
which provides unique opportunity for studying quantum properties in
macroscopic objects.

Generating large quantum superposition of the macroscopic objects is an
essential task in the field of the macroscopic quantum mechanics \cite%
{rmp-84-157,pra-88-033614, rmp-85-471,prl-110-160403}, which provides a good
platform to understand the mechanism of decoherence in macroscopic objects
\cite{ajp-31-6,zpb-59-223}, to check the scope of application of quantum
theory \cite{Weinberg} and to observe the transition between quantum and
classical physics \cite{pra-84-012121}. To this end, the large quantum
superposition, like the Schr\"{o}dinger cat state (simply called 'cat state'
in the following), has been realized in various systems, such as trapped
ions \cite{pra-58-65}, photons \cite{Nature-448-784}, superconducting qubits
\cite{Nature-406-43}, macroscopic current \cite{science-290-773}, and NAMR
\cite{Nature-464-697}. However, those schemes are based on the dynamical
evolution which is sensitive to quantum fluctuations and definitely
unsuitable for preparing macroscopic quantum states. Therefore, it is
desirable to find a robust way to creating the cat state against the quantum
fluctuation.

On the other hand, the geometric phase is only determined by the path of the
state evolution, rather than the initial state distribution or any details
of the path \cite{20,22,zhang,xue}. In this way, operations and processes
based on the geometric phases are robust to the fluctuation and some other
imperfections in the evolution \cite%
{pra-58-65,Nature-448-784,Nature-406-43,Nature-464-697,njp-15-043025}. So
far there are two kinds of geometric phases. The one taking no dynamical
phase is called conventional geometric phase \cite{31}, where removing the
dynamical phase from the evolution is essential to the related operations. As a result,
although the geometric phase is intrinsically resistant to parameter
fluctuation, the additional operations for eliminating the dynamical phases
usually bring in unexpected errors. In contrast, the so-called
unconventional geometric phase \cite{20} remains a non-zero dynamical phase
proportional to the geometric phase by a constant independent of the
parameters. As a result, despite involvement of the dynamical phase, all
geometric advantages are still possessed in the processing with
unconventional geometric phase.

In this work, we demonstrate the possibility to generate the cat state by
displacement operators from the unconventional geometric phase gate (defined
later) in the hybrid system consisting of a charge qubit\cite{rmp-85-623} and a NAMR.
Although we have realized that the unconventional geometric phase has been
achieved in the system of trapped ions \cite{30}, which is analogous to the
model under our consideration, the method working there is not applicable to
the NAMR due to much bigger mass of the NAMR making the coupling to the
charge qubit negligible. However, it is worth pointing out that the
unconventional geometric phase is not robust against the qubit decay, i.e., the charge noise. To
overcome this problem, we consider a transmon-type charge qubit in our work, which
is insensitive to the charge noise due to a large ratio of the Josephson energy
to the charging energy \cite{pra-76-042319}. The decay time
of this kind of charge qubit can reach 0.1 ms experimentally \cite{prb-86-100506}.

Different from the previous schemes involving the unconventional geometric
phase \cite{20,30,zhang,xue,23,24} for quantum computation, our scheme
focuses on generating the cat states based on the displacement operators in
a geometric fashion, which guarantees the robustness of the cat states of the NAMR. 
Moreover, in comparison with a recent proposal \cite%
{pra-88-033614} in which the cat state is created in an optomechanical
cavity, the cat state generated in our hybrid system composed of a charge
qubit and a NAMR can be more 'macroscopic'. In addition, there is a
controllable coupling between the NAMR and the charge qubit, which can be
adjusted by the bias voltage on the NAMR. Therefore, the NAMR and the charge
qubit can be decoupled after the cat state is generated. As a practical
aspect, we suppose that the NAMR owns a very high frequency, which ensures
the NAMR remaining in the ground state in the work temperature of the charge
qubit. Assisted by the operations based on the unconventional geometric
phase with a transmon-type charge qubit, our scheme is robust against the fluctuations from the external light
field within the coherence time, and very promising for achievement using currently available
techniques.

%And the macroscopic NAMR takes a
%larger mass and higher frequency. Moreover, the Hamiltonian in our
%scheme can be used to generate a pure unconventional geometric
%phase and coherent state with a displacement operator, and the
%average phonon number in coherent state takes a larger phonon number. As a result,
%the Schr\"{o}dinger cat cat state in our scheme can be closer to the macroscopic quantum
%state and object.Furthermore, different from the previous schemes, the cat state based on
%the dynamical evolution is in this scheme is sensitive to the parameter fluctuation
%our calculation shows this Schr\"{o}dinger cat state can be .  and guarantee this Schr\"{o}dinger cat
%state based on can be realized with the current experiment setups.

The paper is organized as follows. In the next section, we introduce the
theoretical model and derive the effective Hamiltonian. In Sec. \ref{sec3}, we
review the implementation of an unconventional geometric phase gate.
Generation of the cat states based on the unconventional geometric phase is
discussed in Sec. \ref{sec4}, and we justify the robustness of our scheme in
Sec. \ref{sec5}. A brief conclusion is given in Sec. \ref{sec6}.

\section{Theoretical Model and the effective Hamiltonian}

\label{sec2}

\begin{figure}[htbp]
\includegraphics[width=8cm]{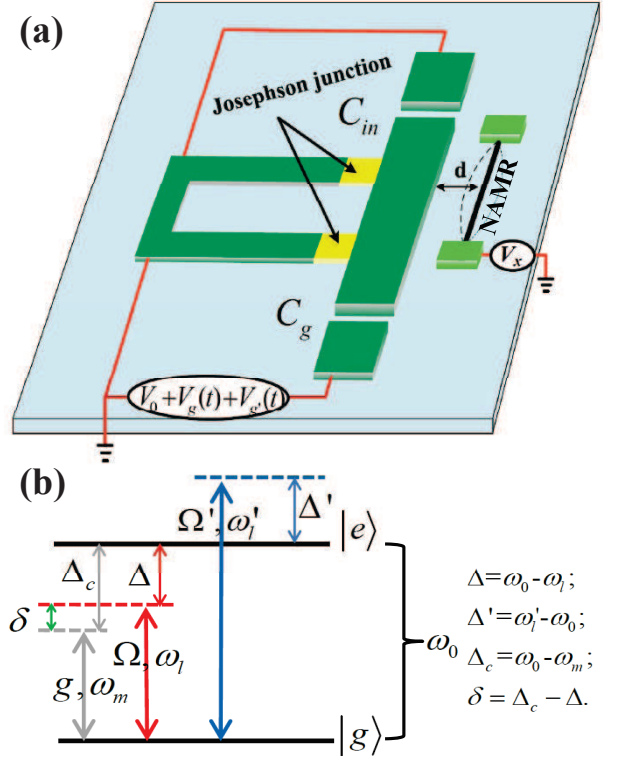}
\caption{(a) Schematic diagram of a NAMR (the black thick line on the right-hand side)
coupled to a transmon-type charge qubit (on the left-hand side) \protect\cite%
{pra-82-032101,prl-92-097204,pra-76-042319}, which could be controlled by an external magnetic field.
Two black dashed curves around the NAMR
present the NAMR in oscillation. Moreover, there is a bias voltage $V_x$
applied on our model, which can be used to control the coupling between the
NAMR and the charge qubit. (b) The corresponding energy levels and
transitions in our scheme, where the parameters are defined in the text.}
\label{scheme}
\end{figure}

As shown in Figure~\ref{scheme}, a metallic NAMR which takes mass $m$,
frequency $\omega _{m}$ and length $L$ is coupled to a superconducting
charge qubit with effective Josephson energy $E_{J}$ and junction capacitance $C_{J}$
by the capacitor $C(x)$ after a static voltage $V_{x}$ is applied. To
suppress the charge noise of the charge qubit, we suppose that the qubit
is well isolated from the rest circuitry by an additional capacitance $C_{in}$, forming
a transmon-type charge qubit \cite{pra-76-042319}. The capacitor $C(x)$ depends on the displacement $x$ of the NAMR
around its equilibrium position. The charge qubit is not only controlled by a DC
voltage $V_{0}$, but also driven by two AC voltages $V_{g}(t)=V\cos
\omega_{l}t$ and $V_{g}^{\prime}(t)=V^{\prime}\cos\omega_{l}^{\prime}t$
via the gate capacitor $C_{g}$. If the charge qubit works in a near
optimal point, which means $[C_{g}V_{0}+C(x)|_{x=0}V_{x}]/2e\approx 0.5$, and the charge energy $E_{c}$
is much less than the effective Josephson energy $E_{J}$ \cite{pra-76-042319,nature-445-515},
the charge noise can be effectively suppressed,
and the transition frequency between the first excited and ground states is $
\omega _{0}\approx \sqrt{8E_{J}E_{c}}/\hbar$ \cite{pra-82-032101}.
By neglecting the higher-order terms regarding the multi-photon excitations, the
Hamiltonian describing our model in the rotating-wave approximation is given
by (in units of $\hbar =1$) \cite{pra-82-032101,prl-92-097204}
\begin{equation}
\begin{array}{rcl}
\hat{H} & = & \frac{1}{2}\omega _{0}\sigma _{z}+\omega _{m}a^{\dag
}a+g(a\sigma ^{+}+a^{\dag }\sigma ^{-}) \\
& + & [(\dfrac{\Omega }{2}e^{-i\omega _{l}t}+\dfrac{\Omega ^{\prime }}{2}
e^{-i\omega _{l}^{\prime }t})\sigma ^{+}+H.c.],
\end{array}
\label{eq001}
\end{equation}
where $\sigma _{z}=\left\vert e\right\rangle \left\langle e\right\vert
-\left\vert g\right\rangle \left\langle g\right\vert $, $\sigma
^{+}=\left\vert e\right\rangle \left\langle g\right\vert $, and $\sigma
^{-}=\left\vert g\right\rangle \left\langle e\right\vert $ with the excited
and ground states $\left\vert e\right\rangle $ and $\left\vert
g\right\rangle $ in the charge qubit, respectively. The operator $a$ ($%
a^{\dag }$), which is the annihilation (creation) operator of the NAMR, can
be written as
\begin{equation}
\begin{array}{rcl}
a & = & \sqrt{\dfrac{m\omega _{m}}{2}}(x+\dfrac{i}{m\omega _{m}}p), \\
a^{\dag } & = & \sqrt{\dfrac{m\omega _{m}}{2}}(x-\dfrac{i}{m\omega _{m}}p),%
\end{array}%
\end{equation}%
with $p$ being the momentum operator of the NAMR. In Eq. (\ref{eq001}), the
first two terms describe the free Hamiltonians of the charge qubit and the
NAMR, respectively. The third term shows the interaction relevant to the
bias voltage $V_{x}$, i.e., a capacitive coupling between the NAMR and the
charge qubit, with the strength
\begin{equation}
g=\frac{4E_{c}N_{x}X_{0}}{d}
\end{equation}%
governed by the charging energy $E_{c}$, the Copper pair number $%
N_{x}=C(0)V_{x} $, the zero-point motion amplitude of the NAMR $X_{0}=1/\sqrt{%
2m\omega _{m}}$, and the distance between the charge qubit and the NAMR $d$.
The last two terms describe the qubit driven by two classical fields with
the frequencies $\omega _{l}$ and $\omega _{l}^{\prime }$, and the Rabi
frequencies $\Omega $ and $\Omega ^{\prime }$, respectively.

In the interaction picture, the Hamiltonian can be rewritten as
\begin{equation}
\begin{array}{rcl}
\hat{H}_{I} & = & (gae^{i\Delta_{c}t}+\frac{\Omega}{2}e^{i\Delta t}+\frac{%
\Omega^{\prime}}{2}e^{-i\Delta^{\prime}t})\sigma^{+}+H.c.,%
\end{array}
\label{eq1}
\end{equation}
with the detunings for the phonon mode of the NAMR and the classical fields
from the charge qubit, respectively, i.e., $\Delta_{c}=\omega_{0}-\omega_{m}$%
, $\Delta =\omega_{0}-\omega _{l}$, and $\Delta^{\prime}=-(\omega_{0}-%
\omega_{l}^{\prime})$ [see Figure \ref{scheme}(b)].

\begin{figure}[tbph]
\includegraphics[width=8cm]{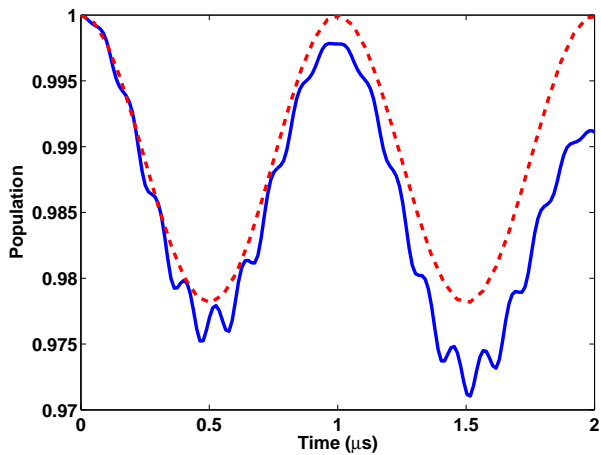}
\caption{(Color online) Time evolution of the population of the state $%
(|g\rangle+|e\rangle)/\protect\sqrt{2}$, where the states for the charge
qubit and the nanomechanical mode are initially in the superposition state $%
(|g\rangle+|e\rangle)/\protect\sqrt{2}$ and in a vacuum state $|0\rangle$,
respectively. Other parameters are $g=2\protect\pi\times3$ MHz, $\Omega=\Omega
^{\prime}=2%
\protect\pi\times30$ MHz, $\Delta_{c}=2\protect\pi\times302$ MHz, and $%
\Delta=\Delta^{\prime}=2\protect\pi\times300$ MHz. The red dashed and blue solid curves are
plotted for comparison, by the Hamiltonians in Eq. (\protect\ref{eq2}) and
Eq. (\protect\ref{eq1}), respectively.}
\label{verify}
\end{figure}

To generate the entanglement state for the NAMR cat state based on the displacement
operators, we derive below
an effective Hamiltonian by employing the method proposed in Refs. \cite%
{zhang,24,28,2801} under the following conditions: (1) $|\Omega |=|\Omega
^{\prime}|$; (2) $\Delta =\Delta ^{\prime} >0$; (3) $\Delta_c>\Delta \gg \{
|g|,|\Omega|, |\Omega ^{\prime}|$\}; (4) $|\Omega |\gg |g|$. The first
condition together with the second condition can completely cancel the Stark
shifts caused by the two driving fields and other related terms. Under the
large detuning condition (3), the charge qubit, initially in the ground
(excited) state, only virtually exchanges phonons with the field of the NAMR
if the operation time is fully within the decoherence time. Moreover, the
last condition ensures that the terms proportional to $|g|^{2}$ are
negligible. Therefore, we obtain the following effective Hamiltonian
\begin{equation}
\hat{H}_{eff}=(\lambda ae^{i\delta t}+\lambda ^{\ast }a^{\dag }e^{-i\delta
t})\sigma_{z},  \label{eq2}
\end{equation}
with $\delta=\Delta_c-\Delta\ll\Delta$ and $\lambda =\frac{\Omega g}{4}(%
\frac{1}{\Delta }+\frac{1}{\Delta _{c}})$. As demonstrated in Figure \ref%
{verify}, the effective Hamiltonian (\ref{eq2}) presents the main
feature of the interaction Hamiltonian (\ref{eq1}), particularly within
the decoherence times of the transmon-type charge qubit and the NAMR (i.e., $t<50$ $\mu$s \cite%
{Nature-464-697,pra-82-032101,prb-86-100506,QIP-8-55}).

\section{The total displacement operator and the unconventional geometric
phase}

\label{sec3}

We present below how to achieve the total displacement operator and generate
the unconventional geometric phase in the NAMR-qubit system using the
effective Hamiltonian (\ref{eq2}). As a starting point, we describe the
derivation of the unconventional geometric phase based on the displacement
operator along an arbitrary path in the phase space \cite{20,22,23,24,30}. For a
displacement operator of the bosonic field,
\begin{equation}
D(\alpha )=e^{\alpha a^{\dag }-\alpha ^{\ast }a},  \label{eq0101}
\end{equation}
where $\alpha$ is the time-dependent displacement parameter for the
creation operator $a^{\dag }$ of the
bosonic field (i.e., the phonon mode of the NAMR in this scheme),
respectively. After experiencing a path which is divided in the phase space into
$N$ short straight sections $\Delta\alpha_{m}$ ($m=1, 2, 3, \cdots$), the
total displacement operation is given by \cite{20,22},
\begin{equation}
\begin{array}{rcl}
D_{t} & = & D(\Delta \alpha _{N})...D(\Delta \alpha _{1}) \\
& = & \exp (iIm\{\sum_{m=2}^{N}\Delta \alpha _{m}\sum_{k=1}^{m-1}\Delta
\alpha _{k}^{\ast }\})D(\sum_{m=1}^{N}\Delta \alpha _{m})\mathbf{,}%
\end{array}
\label{eq103}
\end{equation}
where the Baker-Campbell-Hausdorff formula has been employed. For $%
N\longrightarrow \infty$, the total displacement operator is reduced to
\begin{equation}
D_{t}=e^{i\Theta }D(\int_{\gamma }d\alpha ),  \label{eq0104}
\end{equation}
with the total unconventional geometric phase
\begin{equation}
\Theta =Im\{\int_{\gamma }\alpha ^{\ast }d\alpha \}.
\end{equation}
Different from the conventional geometric phase which removes the dynamical
phase, this total path includes both the geometric phase and the nonzero
dynamical phase with the dynamical phase proportional to the geometric one
\cite{20}. As a result, this total phase $\Theta$ is an unconventional
geometric phase \cite{20}.

When the path is closed, the total displacement operator is rewritten as
\begin{equation}
D_{t}=D(0)e^{i\Theta},  \label{eq0106}
\end{equation}
where the total phase $\Theta$ is only determined by the area of the loop in
the phase space, rather than other factors, such as the quantized state of
the bosonic mode \cite{22}.

According to the definition of the unconventional geometric phase, in the
infinitesimal interval $[t,t+dt]$, the system governed by the effective
Hamiltonian Eq.~(\ref{eq2}) for different states of the charge qubit evolves
as follows,
\begin{equation}
\left\{
\begin{array}{rcl}
|g\rangle |\theta _{g}(t)\rangle & \rightarrow & e^{-i H_{eff}dt}|g\rangle
|\theta _{g}(t)\rangle =D(d\alpha _{g})|g\rangle |\theta _{g}(t)\rangle , \\
|e\rangle |\theta _{e}(t)\rangle & \rightarrow & D(d\alpha _{e})|e\rangle
|\theta _{e}(t)\rangle ,%
\end{array}%
\right.  \label{eq101}
\end{equation}
where $d\alpha_{g}=i\lambda^{\ast }e^{-i\delta t}dt$, $d\alpha
_{e}=-i\lambda ^{\ast }e^{-i\delta t}dt$, and $|\theta _{u}(t)\rangle$ $%
(u=g,e)$ is the state of the nanomechanical phonon mode, which depends on
the qubit state $|u\rangle$ at the time $t$.

If the NAMR is initially in the vacuum state $|0\rangle$, and the charge
qubit is prepared in its ground (excited) state $|g\rangle $ ($|e\rangle $),
after an interaction time $t$, the evolution of the system takes the form of
\begin{equation}
\begin{array}{rcl}
|g\rangle |0\rangle & \rightarrow & e^{i\phi _{g}}D(\alpha _{g})|g\rangle
|0\rangle , \\
|e\rangle |0\rangle & \rightarrow & e^{i\phi _{e}}D(\alpha _{e})|e\rangle
|0\rangle ,%
\end{array}
\label{eq105}
\end{equation}
with
\begin{equation}
\left\{
\begin{array}{rcl}
\alpha _{g} & = & i\int\limits_{0}^{t}\lambda ^{\ast }e^{-i\delta t}dt=-
\frac{\lambda ^{\ast }}{\delta }(e^{-i\delta \tau }-1), \\
\alpha _{e} & = & -i\int\limits_{0}^{t}\lambda ^{\ast }e^{-i\delta
t}dt=-\alpha _{g},%
\end{array}%
\right.  \label{eq106}
\end{equation}
and
\begin{equation}
\left\{
\begin{array}{rcl}
\phi _{g} & = & Im(\int \alpha _{g}^{\ast }d\alpha _{g})\mathbf{=}-\frac{
|\lambda |^{2}}{\delta }(t-\frac{\sin (\delta t)}{\delta }), \\
\phi _{e} & = & Im(\int \alpha _{e}^{\ast }d\alpha _{e})=\phi _{g},%
\end{array}
\right.  \label{eq107}
\end{equation}

We define the operations in Eq. (\ref{eq105}) as unconventional geometric
phase gate. Eq. (\ref{eq106}) shows that, under the condition of $t=2l\pi
/\delta$ ($l=1, 2, 3, \cdots$), the displacement parameter $\alpha_{u}$ for
the state $|u\rangle$ moves along a closed path and returns to the original
point in the phase space of the coherent state $|\alpha _{u}\rangle$.
%{\color{red} The system acquires different unconventional geometric phases, as in Eq. (\ref{eq107}), conditional upon the
%different states of the charge qubit}.
Within a definite period of time, e.g. from $t=2l\pi/\delta$ to $%
t=2(l+1)\pi/\delta$, the state regarding the phonon mode of the NAMR first
evolves from a vacuum state to a coherence state, and then evolves back to
the vacuum state again. The generated unconventional geometric phase can be
controlled by adjusting the detuning $\delta$ [see Eq. (\ref{eq2})].

We would like to point out that the above geometric phases related to the
excited and ground states are the same [see Eq.~(14)]. This is trivial since
it is a global phase. To get the non-trivial unconventional geometric phase,
one can introduce a third level as an auxiliary \cite{zhang}, which is not
governed by Hamiltonian (\ref{eq2}) and can be used for quantum computation
by means of this non-trivial unconventional geometric phase \cite%
{20,30,xue,zhang,23,24}.

In our case here, however, we did not induce such a third level and thus the
involved unconventional geometric phases are identical for the two levels of
the charge qubit. This is because we just focus here on generating entangled
cat states between the NAMR and charge qubit by means of the state-dependent 
displacement operation in a geometric
fashion, rather than based on the phases. Nevertheless, the unconventional
geometric phase gate operation guarantees the robustness of the entangled cat state for
the NAMR and charge qubit in the next section. That results from
the fact that the parameter fluctuations can be absorbed into the
global phase factor by the unconventional geometric phase gate operations.

\section{The cat state with the state-dependent displacement operator}

\label{sec4}

We show specifically how to generate the entangled cat states of the NAMR by the effective
Hamiltonian (\ref{eq2}) using the state-dependent displacement operator
by the unconventional geometric phase gate described in last section.

Suppose that the initial state of the NAMR is in the vacuum state $|0\rangle
$, and the charge qubit is initially in the superposition state $(|g\rangle
+|e\rangle )/\sqrt{2}$. The evolution of the system governed by Eq. (\ref%
{eq105}) is given by
\begin{equation}
\begin{array}{rcl}
\dfrac{|g\rangle +|e\rangle }{\sqrt{2}}|0\rangle & \rightarrow & \dfrac{%
e^{i\phi _{g}}D(\alpha _{g})|g\rangle +e^{i\phi _{e}}D(\alpha _{e})|e\rangle
}{\sqrt{2}}|0\rangle\\
& = & \dfrac{e^{i\phi _{g}}|g\rangle |\alpha _{g}\rangle +e^{i\phi
_{e}}|e\rangle |\alpha _{e}\rangle }{\sqrt{2}}\\
& = & \dfrac{e^{i\phi _{g}}}{\sqrt{2}}(|g\rangle |\alpha _{g}\rangle
+|e\rangle |-\alpha _{g}\rangle ),%
\end{array}
\label{eq205}
\end{equation}%
which is the entangled state between the NAMR and the charge qubit.
In this case, a large enough displacement parameter $\alpha_{u}$
(e.g., $|\alpha_{u}|\gg 1$) for the state $|u\rangle$ guarantees the generated
coherent state with a large phonon number, while the geometric phase
produced only works as a global phase. Nevertheless, the unconventional
geometric phase gate makes sure the robustness in the generation of the entanglement
regarding the NAMR cat state.

According to Eqs. (\ref{eq106}) and (\ref{eq107}), the above entangled state (\ref%
{eq205}) at $t=\pi /2\delta $ is reduced to
\begin{equation}
\begin{array}{rcl}
\dfrac{|g\rangle +|e\rangle }{\sqrt{2}}|0\rangle & \rightarrow & \dfrac{%
|g\rangle |\alpha \rangle +|e\rangle |-\alpha \rangle }{\sqrt{2}},%
\end{array}
\label{eq206}
\end{equation}%
where $\alpha =\frac{2\lambda ^{\ast }}{\delta }=\frac{\Omega g}{2\delta }(%
\frac{1}{\Delta }+\frac{1}{\Delta +\delta })$ and the global phase has been
ignored. The definition of the parameter $\alpha $ shows that the average
phonon number in the coherent state can be changed if we adjust the detuning
$\delta $. Therefore, it is possible to get very large phonon coherent states 
in the entangled state which
takes a parameter $\alpha \simeq \frac{\Omega g}{\delta \Delta }$ with $%
\delta \ll |\lambda |$ and $\delta \ll \Delta $ in the ideal condition.

In addition, once we turn off the bias voltage after finishing the
operations above, the NAMR and the charge qubit are decoupled. By a single
qubit operation, i.e., $|e\rangle \rightarrow (|g\rangle -|e\rangle )/\sqrt{2%
}$ and $|g\rangle \rightarrow (|g\rangle +|e\rangle )/\sqrt{2}$ \cite%
{QIP-8-55}, on the charge qubit, Eq. (\ref{eq206}) changes as
\begin{equation}
\begin{array}{lll}
\dfrac{|g\rangle |\alpha \rangle +|e\rangle |-\alpha \rangle }{\sqrt{2}} &
\rightarrow & \dfrac{|g\rangle (|\alpha \rangle +|-\alpha \rangle
)+|e\rangle (|\alpha \rangle -|-\alpha \rangle )}{2}.
\end{array}
\label{eq207}
\end{equation}
After the measurement is performed on $|g\rangle$ and $|e\rangle$ of the charge qubit,
the final states in Eq.~(\ref{eq207}) collapse into the cat states
$(|\alpha \rangle +|-\alpha\rangle )/\sqrt{2}$ and $(|\alpha \rangle -|-\alpha \rangle )/\sqrt{2}$ of the NAMR,
respectively, which can be detected by
applying a static magnetic field and an alternating current \cite
{pra-82-032101}.

\section{Simulation and Discussion}

\label{sec5}

We estimate numerically how well our scheme works under noisy environment.
For convenience of the discussion, we take the entangled cat state in Eq. (\ref{eq206}) as
an example by considering decays from the charge qubit and the NAMR. So the
master equation for our scheme is given by
\begin{equation}
\begin{array}{rcl}
\dot{\rho} & = & -i[H_{I},\rho ]+\frac{\Gamma }{2}(2\sigma \rho \sigma
^{+}-\sigma ^{+}\sigma \rho -\rho \sigma ^{+}\sigma ) \\
&  & +\frac{\gamma }{2} (2a\rho a^{\dag }-a^{\dag }a\rho -\rho a^{\dag }a),%
\end{array}
\label{eq02221}
\end{equation}
where $\rho$ is the reduced density operator of the system, $\Gamma $ and $%
\gamma$ are the charge qubit decay rate and the NAMR decay rate,
respectively. Note that the thermal phonons of the NAMR have been ignored in
Eq.~(\ref{eq02221}), due to the assumption that the thermal phonons can be
neglected since the high-frequency NAMR can be remained in its ground state
at the work temperature ($T=20$ mK) of the charge qubit \cite{QIP-8-55,
Nature-464-697}.

\begin{figure}[tbph]
\includegraphics[width=8cm]{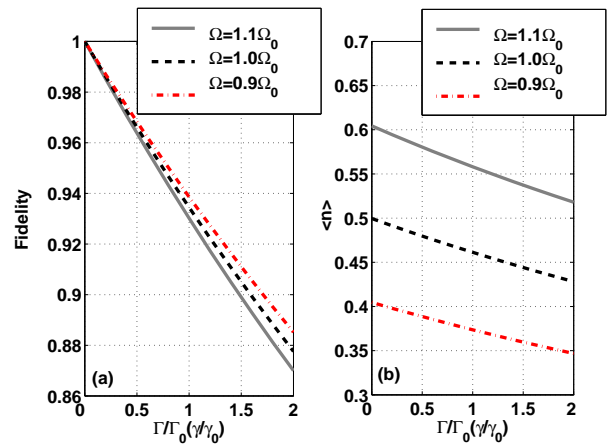}
\caption{(Color online) (a) Numerical simulation of the fidelity of the entangled cat state
 in Eq. (\protect\ref{eq206}) versus the NAMR decay and charge qubit
decay for different Rabi frequencies. (b) The corresponding average phonon number
$Tr(\rho(t)a^{\dag}a)$ in the entangled cat state. $0.9\Omega_0$ and $1.1\Omega_0$
mean $10\%$ and  $-10\%$ fluctuations, respectively. Here for simplicity we assume $\protect
\gamma /\protect\gamma _{0}=\Gamma /\Gamma _{0}$. The parameters take values
$\protect\omega _{m}=2\protect\pi \times 1000$ MHz, $Q=10^{5}$, $g=2\protect
\pi \times 2$ MHz, $\Omega_{0}=2\protect\pi\times 20$ MHz, $\Gamma _{0}=2\protect\pi\times 0.02$
MHz, $\protect\gamma_{0}=\protect\omega_{m}/Q=2\protect\pi\times0.01$ MHz,
$\Delta_{c}=2\protect\pi\times 200.2$ MHz, and $\Delta =2\protect\pi\times 200$ MHz.}
\label{fidelity}
\end{figure}

We define the fidelity of the entangled cat state
\begin{equation}
F=\langle \Psi |\rho (t)|\Psi \rangle ,  \label{eq402}
\end{equation}
where $t$ is the time for producing the entangled cat state, and $|\Psi\rangle$ is a target state relevant to the
initial state. In our scheme, the initial state and the target state are
$(|g\rangle +|e\rangle )|0\rangle /\sqrt{2}$ and $(|g\rangle |\alpha \rangle
+|e\rangle |-\alpha \rangle )/\sqrt{2}$, respectively. Using the realistic
parameter values from Refs. \cite
{Nature-464-697,prl-93-070501,nature-445-515,QIP-8-55,prb-70-205304,
prb-69-125339,pra-82-032101}, we show in Figure \ref{fidelity} the numerical
results for the fidelities of the generated state and the corresponding average phonon number $Tr(\rho(t)a^{\dag}a)$ versus
the decays of the charge qubit and the NAMR with respect to different Rabi frequencies, where we consider $\pm 10\%$ change of the Rabi
frequency by parameter fluctuations.

Our simulation shows the fidelity for the generated entangled state decreases with
the increase of the decays. Here, we suppose
$\gamma_{0}$=(50 $\mu$s)$^{-1}$ ($\Gamma_{0}$=(100 $\mu$s)$^{-1}$) as the
realistic transmon-type charge qubit (NAMR) decay rate \cite{prb-86-100506, Nature-464-697}.
Figure \ref{fidelity} (a) means that the fidelity of the
entangled cat state is sensitive to the decays of the charge qubit and the NAMR, and the
unconventional geometric phase gate can be robust against the parameter 
fluctuation rather than the qubit decay. The decay of the charge qubit
causes the flip of the charge qubit states and the decay of the NAMR leads
to the phonon leakage from the coherent states.

Moreover, according to the discussion in Section 4, the operation time
for preparing the entangled cat state strongly depends
on the detuning $\delta$ [see Eq. (\ref%
{eq106})]. Provided fixed Rabi frequencies regarding the driving fields and
the NAMR, the phonon coherent state can be very large if $\delta $ is very small.
However, we need a long time to achieve the large coherent states in such a case, which
is usually longer than the decay time. In contrast, if $\delta $ is very
large, the operation time can be very short, but the average phonon number
in the entangled state is very small. As a trade-off, the
detuning $\delta$ must be chosen carefully to be small enough but make sure
the operation to be finished within the decoherence time of the system. Considering
the parameters in Figure \ref{fidelity}, we may set the operation time $t=1.25$ $\mu$s,
and then we can find that the lowest average phonon number of the entangled cat state 
in our simulations is $\left\langle n\right\rangle =0.347$ (See the lowest point in 
red dashed-dotted line in Figure \ref{fidelity} (b)). In the absence of decoherence 
and parameter fluctuations, the corresponding maximal value of 
$\left\langle n\right\rangle$ in Figure \ref{fidelity} can reach 1.0, which
is larger than $\left\langle n\right\rangle =0.1$ proposed in Ref. \cite{pra-88-033614}.

In addition, there are some other factors affecting the fidelity of the generated
entangled cat state under the noisy environment.
These factors include
the uncertainties and fluctuations of the parameters, which can change the average
phonon number in the coherent state. As an example, we have calculated in
Figure \ref{fidelity}(a) the influence due to $10\%$ fluctuation in the Rabi frequency,
which demonstrates the robustness to the fluctuations to some extent.
Particularly, we find that $-10\%$ fluctuation of the Rabi frequency (i.e., $0.9\Omega_0$) corresponds to
a higher fidelity of the entangled cat state, which is due to the fact that the average
phonon number in the coherent states is proportional to the Rabi frequency
$\left\langle n\right\rangle =|\alpha|^{2}\sim|\frac{\Omega g}{\delta\Delta}|^2$,
when the operation is performed within the time for finishing a quarter of a Rabi
oscillation. This implies that the less Rabi frequency leads to less average phonons
in the coherent state (see Figure \ref{fidelity} (b)), which is less sensitive to the decay of the NAMR.
Nevertheless, the simulations in Figure \ref{fidelity} show that the validity of our scheme
is restricted by the charge qubit decay, since the scheme based on the geometric phase is never
robust against the qubit decay.

\section{Conclusion}

\label{sec6}

In conclusion, we have shown the possibility to achieve the cat state of the NAMR in a
NAMR-charge qubit system. The key point of our scheme is the employment of
the state-dependent displacement operator based on the unconventional
geometric phase gate, which makes the cat state generated with a geometric
feature. As a result, our scheme owns some distinct advantages, such as the
robustness against the parameter uncertainties and fluctuations, the
feasibility of the ground state of the NAMR in the work temperature of the
charge qubit, independent control of the charge qubit and the NAMR, and the
possibility of producing large cat states. In addition, our scheme can also
be extended to the Von Neumann measurement based on geometric feature of the operations
\cite{pra-85-022129}. Particularly, we argue that our
scheme can be achieved using currently available techniques.

\section*{Acknowledgments}

JQZ thank Yue Li, Zhi-Jiao Deng, Zheng-Yuan Xue, Zhangqi Yin and Yue-Yue Chen for helpful
discussions. This work was supported by the National Natural Science
Foundation of China (Grants Nos. 11174027, 11121403, 11274352 and 11304366) and the
China Postdoctoral Science Foundation (Grant Nos. 2013M531771 and 2014T70760).

\end{document}